\RequirePackage[2020-02-02]{latexrelease}
\documentclass[aps,prb,twocolumn,amsmath,superscriptaddress,amssymb]{revtex4}
\usepackage{graphicx}
\usepackage{enumitem}
\usepackage[colorlinks,bookmarks=false,citecolor=red,linkcolor=blue,urlcolor=blue]{hyperref}

\newcommand{\nn}{\nonumber}          
\newcommand{\bea}{\begin{eqnarray}}          
\newcommand{\eea}{\end{eqnarray}}

\allowdisplaybreaks

\begin{document}

\title{Scattering off a junction}
\author{Eric Tan}
\email{eric.tian.tan@gmail.com}
\affiliation{University of Western Ontario,  
London, Ontario N6A 3K7, Canada}
\affiliation{Department of Physics, Brock University, St. Catharines, Ontario L2S 3A1, Canada}
\author{R. Ganesh}
\email{r.ganesh@brocku.ca}
\affiliation{Department of Physics, Brock University, St. Catharines, Ontario L2S 3A1, Canada}
\date{\today}

\begin{abstract}
Scattering off a potential is a fundamental problem in quantum physics. It has been studied extensively with amplitudes derived for various potentials. In this article, we explore a setting with no potentials, where scattering occurs off a junction where many wires meet. 
We study this problem using a tight-binding discretization of a star graph geometry -- one incoming wire and $M$ outgoing wires intersecting at a point.  
An incoming wave arrives at the junction and scatters. One part is reflected along the same wire while the rest is transmitted along the others. Remarkably, the reflectance increases monotonically with $M$, i.e., the greater the number of outgoing channels, the more the particle bounces back. In the $M \rightarrow \infty$ limit, the wave is entirely reflected back along the incoming wire. We rationalize this observation by establishing a quantitative mapping between a junction and an on-site potential. To each junction, we assign an equivalent potential that produces the same reflectance. As the number of wires ($M$) increases, the equivalent potential also increases. A recent article by one of us has drawn an equivalence between junctions and potentials from the point of view of bound state formation. 
Our results here show that the same equivalence also holds for scattering amplitudes.
We verify our analytic results by simulating wavepacket motion through a junction. 
We extend the wavepacket approach to two dimensions where analytic solutions cannot be found. An incoming wave travels on a sheet and scatters off a point where many sheets intersect. 
As in the one-dimensional problem, we assign an equivalent potential to a junction. 
However, unlike in 1D, the equivalent potential is momentum-dependent. 
Nevertheless, for any given momentum, the equivalent potential grows monotonically with the number of intersecting sheets. Our findings can be tested in ultracold atom setups and semiconductor structures. 
\end{abstract}
                                 
\maketitle
\section{Introduction}
The problem of scattering has a long, illustrious history in classical and quantum physics. In its simplest form, a wave travels through free space and reaches a local potential. A part of the wave is then transmitted while the rest is reflected or scattered\cite{Schiff,Sakurai,Goldberger,Newton,Joachain}. Reflection and transmission coefficients have been calculated for various potentials (e.g., see Ref.~\onlinecite{Ahmed2021}). 
In this article, we discuss scattering off a junction and compare it to that off a potential. A junction is similar to a potential in that it breaks translational symmetry. As the strength of a potential can be tuned, so can the `degree of connectivity' at a junction. This quantity represents the number of wires that meet at the junction. Our key result is that in one dimension, a junction and a local potential are quantitatively equivalent.
In particular, as the connectivity of a junction increases, the equivalent potential increases. This leads to a counter-intuitive result: the larger the number of outgoing channels at a junction, the more is the reflection back along the incoming channel.

Junctions have been experimentally studied in various settings. Ultracold atoms have been loaded onto Y-junction and X-junction geometries\cite{Cassettari2000,Dumke2002}. 
Semiconductor structures with T geometries (three wires meeting at a point)\cite{Goni1992,Hasen1997} and X geometries (four wires meeting at a point)\cite{Berggren1991} have been explored. Networks built from arrangements of junctions have also been studied using confined electron gases\cite{Naud2001} and microwaves\cite{Hul2004}. To understand transport in junctions, it is crucial to understand the scattering process where a wavepacket impinges on a junction. This acquires significance in fields such as coherent matter wave physics\cite{Bongs2004,Zimmermann2007} or atomtronics\cite{Amico2022}. Many studies have focussed on complex effects caused by interactions, transverse potentials, etc. In this article, we demonstrate that the simplest version of the problem -- a free particle scattering off a simple junction -- already shows interesting and surprising properties.

From a theoretical point of view, transmission and scattering at junctions have been extensively studied in the field of quantum graphs\cite{Andrade2016}. Typically, the Schr\"odinger equation is solved on each smooth wire. At a junction where several wires meet, wavefunctions are related to one another by a scattering matrix. Unitarity imposes strong constraints on this matrix. For example, in a star graph where $M$ wires meet at a point, the net transition amplitude must scale as $\sim 1/M$\cite{Gratus1994}. In other words, for large $M$, an incoming wave is entirely reflected back along the same wire. 
However, in order to explicitly find the scattering matrix, a boundary condition must be imposed at the junction -- for example, see Ref.~\onlinecite{Andrade2016} and references therein. 
The boundary condition is not fully determined by physical considerations (such as probability conservation).
For example, the Robin-type boundary condition involves a free parameter which cannot be fixed or tuned by any known experimental means. As such, the choice of boundary condition is arbitrary and can bias the outcome. 
In this study, we circumvent this issue by taking a tight binding approach where the space of intersecting wires is discretized. As we show below, this allows for finding scattering coefficients in an unbiased manner. 

The tight binding setup can also be directly realized with ultracold atomic gases\cite{Zakrzewski2007}, superconducting circuits\cite{Weiss2021}, etc. It can handle potentials and junctions on the same footing.
Recently, one of the present authors has studied the formation of bound states at junctions\cite{He2023}. Using a tight binding description, junctions of thin wires were argued to generically host a bound state. The bound state arises from additional kinetic-energy terms involving shuttling among wires in the vicinity of the junction. This phenomenon is also seen in higher dimensions, e.g., when multiple 2D sheets or 3D spaces intersect one another. Bound states are intimately related to the scattering problem. As is well known, bound states can be viewed as poles in the scattering coefficients\cite{Sakurai,Ahmed2021}. In this article, we extend junction-potential equivalence from bound states to scattering amplitudes.

\begin{figure}
\includegraphics[width=3in]{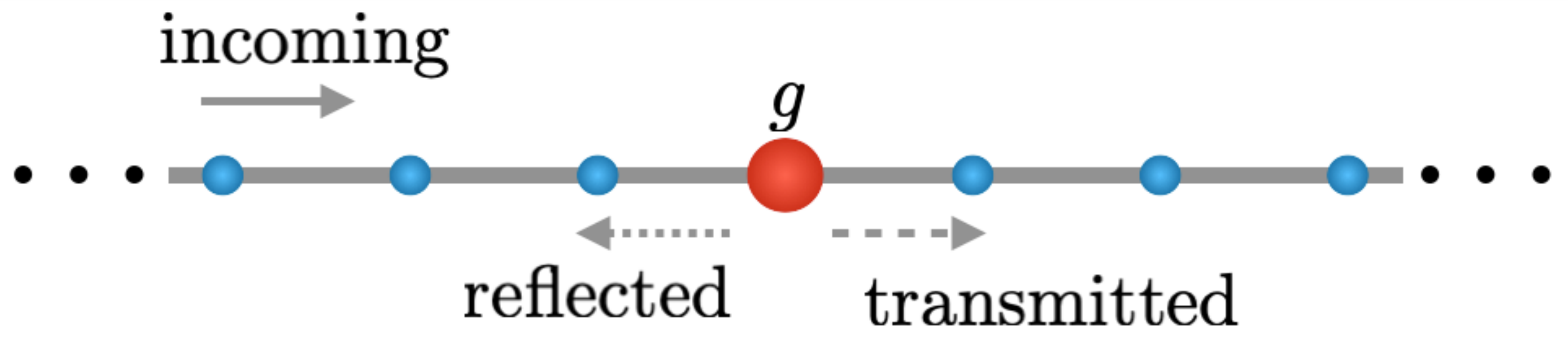}
\caption{Scattering off a potential. A wave travels along a wire, shown as a collection of sites connected by bonds. It encounters an attractive potential at one particular site and scatters. } 
\label{fig.pot}
\end{figure}

\section{Scattering off a potential: tight-binding approach}
\label{sec.potential}
To set the stage, we first discuss scattering off a delta-function potential. While textbooks describe this problem in the continuum using appropriate boundary conditions, we take a tight-binding approach. This will allow us to handle the junction problem on the same footing in later sections.

We consider an infinite wire with sites labelled by integers $j$ as shown in Fig.~\ref{fig.pot}. A particle moves on this wire by `hopping' between nearest neighbours. When the particle is at site $j=0$, it experiences an attractive potential of amplitude $g$. In second quantized notation, this problem is described by the Hamiltonian 
\bea
H = -t \sum_{j} \big[ c_{j+1}^\dagger c_j +c_{j-1}^\dagger c_j  \big] - g~ c_0^\dagger c_0.
\label{eq.tbHampot}
\eea

\subsection{The scattering state}
We seek to find stationary states that encode scattering. We propose the following ansatz for a stationary state,
\bea
\psi_j = \left\{ \begin{array}{c}
A e^{i k j} + B e^{- i k j}, ~~ j< 0, \\
C, ~~j = 0, \\
D e^{i k j}, ~~j> 0.
\end{array}\right.
\label{eq.potansatz}
\eea
Here, $A$ represents the amplitude of the incoming wave, coming in from $-\infty$. The momentum of the wave is denoted as $k$. $B$ and $D$ represent the amplitudes of the reflected and transmitted waves. We assume that this ansatz represents an eigenstate with energy $E$. 

The eigenvalue equation reduces to the following condition on each site, $-t\sum_{m(n)} \psi_m = E \psi_n$. Here, $m(n)$ represents the neighbours of site $n$. For a stationary state, this condition must be satisfied at every site $n$. This leads to three independent relations: one arising from sites to the left of the potential ($n<0$), one at the potential itself and the last arising from sites to the right ($n>0$). These relations allow us to solve for B, C and D in terms of A, as shown in the appendix. We write $B = R (k,g)A$ and $D = T (k,g) A$, where $R$ and $T$ are the reflection and transmission coefficients respectively. We find
\bea
R(k,g) = \left[ \frac{-g}{ g + 2it \sin k} \right]; ~T(k,g) =  \left[ \frac{ 2it \sin k}{g + 2it \sin k} \right].
\label{eq.coeffs}
\eea
The `reflectance' or reflection probability is given by 
\bea
\left| R(k,g) \right|^2 = \frac{\tilde{g}^2}{ \tilde{g}^2 +  \sin^2 k},
\label{eq.refg}
\eea
where $\tilde{g} = g/2t$ is a dimensionless measure of the potential strength.

\subsection{Bound state from scattering coefficients}
\label{ssec.bound}

It is well known that poles of scattering coefficients correspond to bound states\cite{Newton_book}. A bound state occurs when there is no incoming wave, but reflected and transmitted pieces survive -- with exponential decay rather than propagating waves. This can be interpreted as an extension of the wavevector $k$ to complex values, $k \rightarrow i \kappa$. The reflected component would take the form $B e^{\kappa j}$ $(j<0)$ while the transmitted component would be given by $D e^{-\kappa j}$ $(j>0)$. The two forms lead to an exponentially localized solution, provided $\kappa$ is real and positive. To have a non-zero value of B when A vanishes, $R(i\kappa,g)$ must diverge. This corresponds to a pole in the reflection coefficient. In the same manner, a bound state must also correspond to a pole in $T(i\kappa,g)$.

As seen in Eq.~\ref{eq.coeffs} above, $R(i\kappa,g)$ and $T(i\kappa,g)$ share the same denominator. We have a pole when this denominator vanishes. We find a single pole with
\bea
{\kappa} = \ln \{ \tilde{g} +\sqrt{\tilde{g}^2 +1}\}.
\eea
Here, $\kappa$ is a decay constant. When the potential strength $\tilde{g}$ increases, $\kappa$ increases monotonically; the bound state becomes progressively more bound. Note that a pole only occurs if the potential is attractive, i.e., when $g < 0$.

\section{Scattering off a junction}
\label{sec.junction}

\begin{figure}
\includegraphics[width=3in]{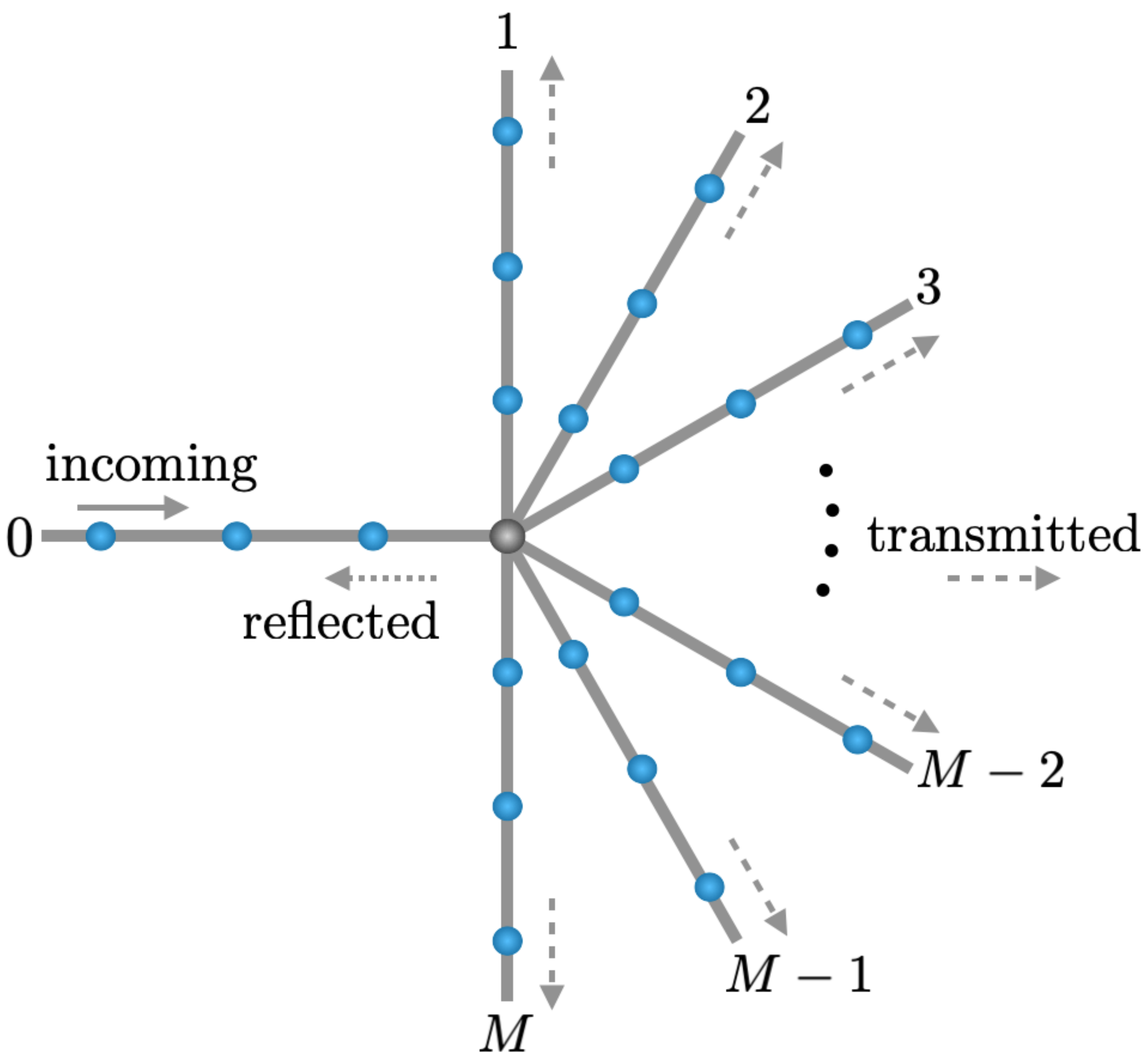}
\caption{Scattering off a junction. A wave travels along the `incoming' wire. It encounters a junction with $M$ outgoing wires and scatters. The incoming wire is labelled as `$0$'. A part is reflected along the same wire, while the rest is transmitted through the remaining $M$ wires. } 
\label{fig.junction}
\end{figure}

We take the junction geometry shown in Fig.~\ref{fig.junction}. We discretize this space with points represented as $(j,\ell)$, where $\ell$ identifies a wire and $j$ refers to points within the wire. We designate the incoming wire as $\ell=0$ and outgoing wires as $\ell =1, 2, \ldots,M$. Sites of the form $(0,\ell)$ are identified, i.e., they represent the same site irrespective of $\ell$. This common site serves as the junction. 
On the outgoing wires, sites are labelled as $j=1,2,\ldots$ as we move away from the junction. On the incoming wire, we label sites as $j=-1,-2,\ldots$ as we move away from the junction.

A particle moving in this geometry is described by a tight binding Hamiltonian,
\bea
 H = H_{\ell=0} + \sum_{\ell = 1}^M H_{\ell} + H_{junction}.
 \label{eq.tbHam}
\eea
The first piece represents hopping on the incoming wire,
\bea
H_{\ell=0} = -t \sum_{j=-1,-2,-3,\ldots}\big[ c_{j,0}^\dagger c_{j-1,0} +  c_{j-1,0}^\dagger c_{j,0} \big].
\eea
The second represents hopping on each outgoing wire,
\bea
H_{\ell \neq 0}=  - t \sum_{j=1,2,3,\ldots} \big[ c_{j,\ell}^\dagger c_{j+1,\ell} +  c_{j+1,\ell}^\dagger c_{j,\ell} \big].
\eea
The third encodes hopping from the junction to each wire,
\bea
\nn H_{junction} = -t \big\{ c_{-1,0}^\dagger c_0 + c_0^\dagger c_{-1,0} 
\big\}\\
-t \sum_{\ell=1,2,\ldots,M} \big\{ c_{1,\ell}^\dagger c_0+c_0^\dagger c_{1,\ell}\big\}. 
\eea
Note that all terms in the Hamiltonian represent kinetic energies of hopping. There are no potential terms. 

\subsection{The scattering state}
\label{ssec.junctionansatz}
To describe a steady state, we propose a scattering ansatz of the form 
\bea
\psi (j,\ell) = \left\{ \begin{array}{c}
A e^{i k j} + B e^{- i k j}, ~~ j< 0, ~\ell = 0 \\
C, ~~j = 0, \\
D e^{i k j}, ~~j> 0 , ~\ell =1,2,\ldots, M.
\end{array} \right.
\label{eq.Mansatz}
\eea
As with the potential problem, $A$ is the amplitude of the incoming wave. $B$ and $D$ are amplitudes of reflected and transmitted waves respectively. We have taken the transmitted wave to be symmetrically distributed among the $M$ outgoing wires. This assumption follows from the symmetry of the junction geometry. It is also in agreement with numerical results presented in Sec.~\ref{sec.wavepacket} below.

We express $B = R (k,M)A$ and $D = T (k,M) A$, where $R$ and $T$ are the reflection and transmission coefficients.
We demand that this ansatz be an eigenstate of the Hamiltonian given in Eq.~\ref{eq.tbHam}. Relegating details to the appendix, we find 
\bea
\nn R(k,M) &=& \Big[ \frac{  (1-M)  e^{ik} 
 }{  M  e^{ik} 
  -  e^{-ik} }\Big],\\
  T(k,M) &=& \Big[
 \frac{ 2 i  \sin k}{ M  e^{ik} 
 -  e^{-ik} }
 \Big].~~~
 \label{eq.MRT}
\eea 
This leads to a reflection probability  
\bea
\left| R(k,M) \right|^2 = \frac{
 f_M^2 
 }{
 f_M^2 +  \sin^2 k 
 }, 
 \label{eq.refM}
\eea
where $f_M= \frac{1}{2}\Big(\sqrt{M} - \frac{1}{\sqrt{M}}\Big)$. 
This quantity is a dimensionless measure of the `degree of connectivity' of the junction. When $M=1$, we have one incoming wire and one outgoing wire; the system forms one smooth wire with no junction. In this limit, $f_M$ vanishes. For $M>1$,  $f_M$ increases monotonically with $M$.

\subsection{Bound state from scattering coefficients}
\label{ssec.bsM}
As in Sec.~\ref{ssec.bound}, bound states can be found from poles of $R(k,M)$ and $T(k,M)$. To allow for exponential decay, we extend the wavevector $k$ to complex values, $k \rightarrow i \kappa$. We then set the denominators in Eq.~\ref{eq.MRT} to zero. 
We find a single bound state with
\bea
\kappa = \frac{\ln M}{2}. 
\eea 
Note that $\kappa$ is positive for any $M > 1$, indicating a bound state. This pole represents a localized state that is peaked at the junction, decaying exponentially along each wire. The decay constant, denoted by $\kappa$, is the same on all wires. As the junction's degree of connectivity ($M$) increases, so does $\kappa$. This indicates that the bound state becomes progressively more bound. This expression for $\kappa$ agrees with the result in Ref.~\onlinecite{Khatua2021} (after replacing $M \rightarrow 2M-1$, to align with the notation therein).

\section{Equivalence between a junction and a potential}
\label{sec.equivalence}

\begin{figure}
\includegraphics[width=3.4in]{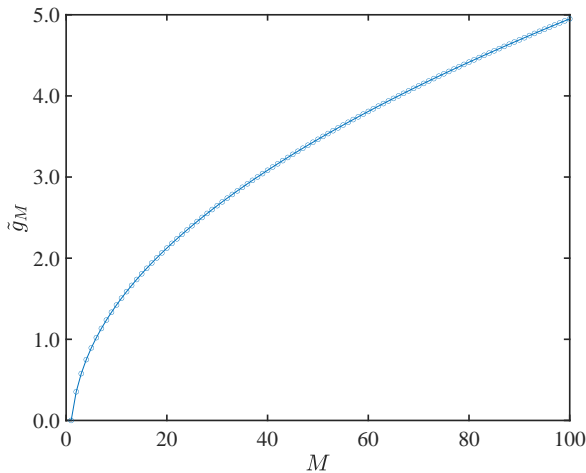}
\caption{Equivalent potential strength vs. degree of connectivity of a junction, $\tilde{g}_{M}$ vs. $M$. } 
\label{fig.eqMforg}
\end{figure}

We have found expressions for reflectance in potential-scattering and junction-scattering. 
Comparing Eqs.~\ref{eq.refg} and \ref{eq.refM}, we find the same functional form. This reveals a quantitative equivalence. A junction with $M$ outgoing wires has the same reflectance as a potential of strength $g$, where $g$ and $M$ satisfy $\tilde{g} = f_M$. Note that $\tilde{g}$ and $f_M$ are monotonic functions of $g$ and $M$ respectively.
This relation is the same as the equivalence condition derived in Ref.~\onlinecite{He2023} on the basis of bound state properties. Here, we have demonstrated that the same equivalence holds for scattering (reflectance) as well. 
In Fig.~\ref{fig.eqMforg}, we plot the equivalent potential strength ($\tilde{g}_M$) vs. the degree of connectivity of a junction, $M$.

The equivalence here is based on reflectance. A potential and a junction are considered equivalent if they produce the same reflection probability. However, the phase of the reflected wave may be different in the two cases.  
For example, when $k=\pi/2$, $R(k=\pi/2,M)$ is real for any $M$. However, $R(k=\pi/2,g)$ carries a non-zero phase. This difference can have observable consequences in the case of a multichromatic incoming wave (with contributions from many frequencies). The junction and its equivalent potential will reflect each frequency component with a different phase shift. As a result, the reflected wavefront may look very different in the two cases.

In the case of a potential, it is easy to understand that reflectance increases with potential strength. In the limit of a very large potential ($g/t \rightarrow \infty$), the reflectance reaches unity -- an incoming particle is entirely reflected, with no transmission across the potential. The equivalence discussed above leads to a surprising observation for junctions. As the degree of connectivity ($M$) increases, so does the reflectance. For a junction with two outgoing channels, the minimum reflection probability is $\sim$ 11\% (from Eq.~\ref{eq.refg} with $k\rightarrow \pi/2$). When the number of outgoing channels increases to ten, minimum reflectance reaches $\sim$ 67\%. With a hundred outgoing channels, this quantity touches $\sim$ 96\%. For any $M >1$, the maximum reflection probability is unity which occurs in the long wavelength limit ($k\rightarrow 0$), as seen from Eq.~\ref{eq.refg}.

\section{Wavepacket scattering}
\label{sec.wavepacket}
Secs.~\ref{sec.potential} and \ref{sec.junction} discuss scattering from a steady-state perspective. Reflectance and transmittance were obtained from a scattering-eigenstate of the Hamiltonian. We now discuss scattering from a dynamical perspective with wavepackets that impinge on a potential or a junction. We take a numerical tight binding approach following Ref.~\onlinecite{Staelens2021}. We begin with the geometry of Fig.~\ref{fig.junction}, with wires consisting of $L$ sites. Each wire, at one end, is connected to a potential or a junction. At the other end, we have an open boundary. The open end does not affect our results as the wavepacket does not reach the boundary during the course of the simulation.  

We construct a Gaussian wavepacket that serves as the initial wavefunction,
\bea
\psi_{j,\ell=0} = \frac{1}{(2\pi \alpha^2)^{1/4}} \exp\Big\{ 
-\frac{1}{4} \frac{(j-j_0)^2}{\alpha^2}
+ i k_0(j-j_0)
\Big\}.~~ 
\eea
The entire weight lies on the incoming wire, with zero amplitude on outgoing wire(s). The amplitude is peaked at site $j_0$ and spread on either side over a width set by $\alpha$. The parameter $k_0$ encodes momentum of the wavepacket. When $k_0 =0$, we expect the wavefunction to remain centred at $j_0$, while spreading on either side with time. With a positive value of $k_0$, the wavepacket will move towards the potential/junction. As shown in Ref.~\onlinecite{Staelens2021}, $k_0 = \pi/2$ is an ideal choice for simulations. It leads to minimal spread of the wavefunction, with coherence retained over a relatively long time scale. 

Starting with this initial wavefunction, we time-evolve using the tight binding Hamiltonian. We take the Hamiltonian of Eq.~\ref{eq.tbHampot} for potential scattering and that of Eq.~\ref{eq.tbHam} for junction scattering. We have performed simulations with various values of $\alpha$ (initial width), $k_0$ (initial velocity), $g$ (potential strength) and $M$ (degree of connectivity at a junction). We obtain consistent results that are in line with the steady-state arguments in previous sections. 

\begin{figure*}
\includegraphics[width=7in]{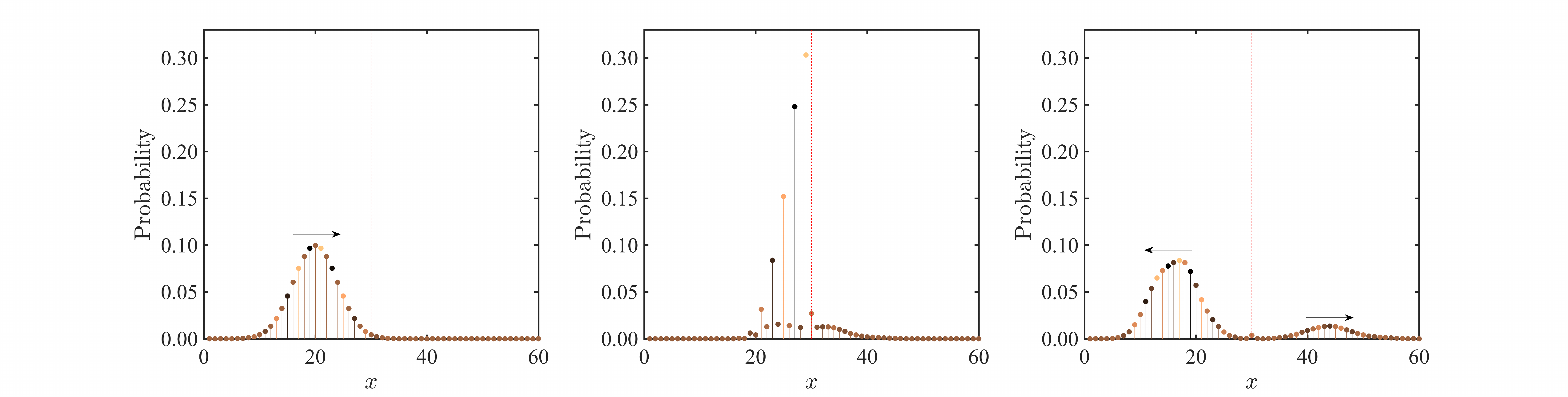}
\caption{Scattering off a potential. We have a 60-site wire, with sites labelled by $x \in [1,60]$. An incoming wavepacket moves towards increasing $x$ (left) and scatters off a potential at $x=30$ (centre). Subsequently, a reflected wavepacket moves left while a transmitted component moves to the right (right). In each plot, the wavefunction amplitude is shown as height, while the marker colour represents phase. As the phase is $k_0=\pi/2$, we see that the phase varies with a period of four sites. } 
\label{fig.scatpot}
\end{figure*}

Fig.~\ref{fig.scatpot} shows the evolution of the wavepacket as it impinges on a potential. After the scattering process, we are left with two propagating wavepackets -- one moving back along the incoming wire and one moving forward on the outgoing wire. We identify these as reflected and transmitted components. Reflectance or reflection probability is simply the weight of the reflected component. As we show below, the reflectance grows monotonically with potential strength.

\begin{figure*}
\includegraphics[width=3.5in]{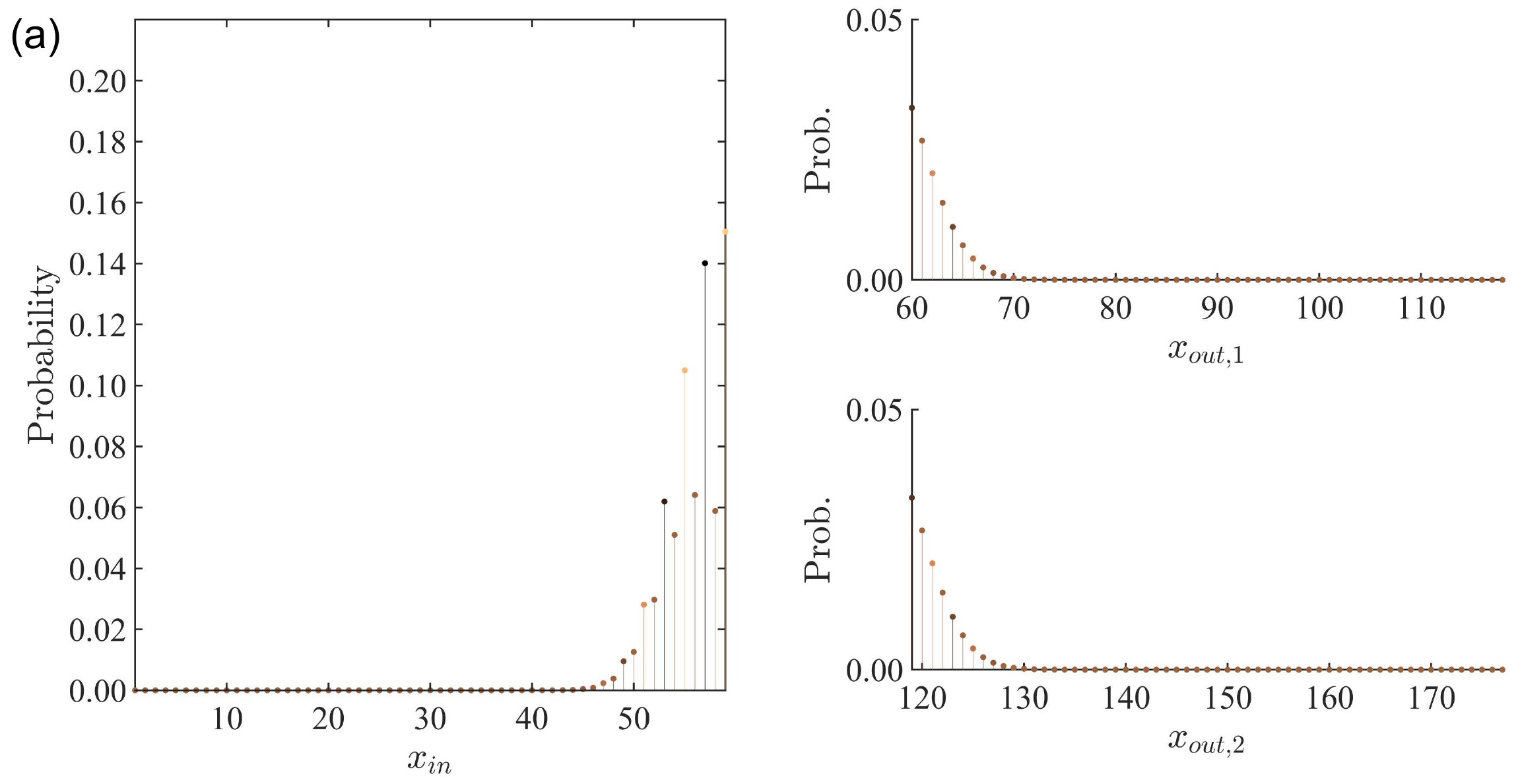}
\includegraphics[width=3.5in]{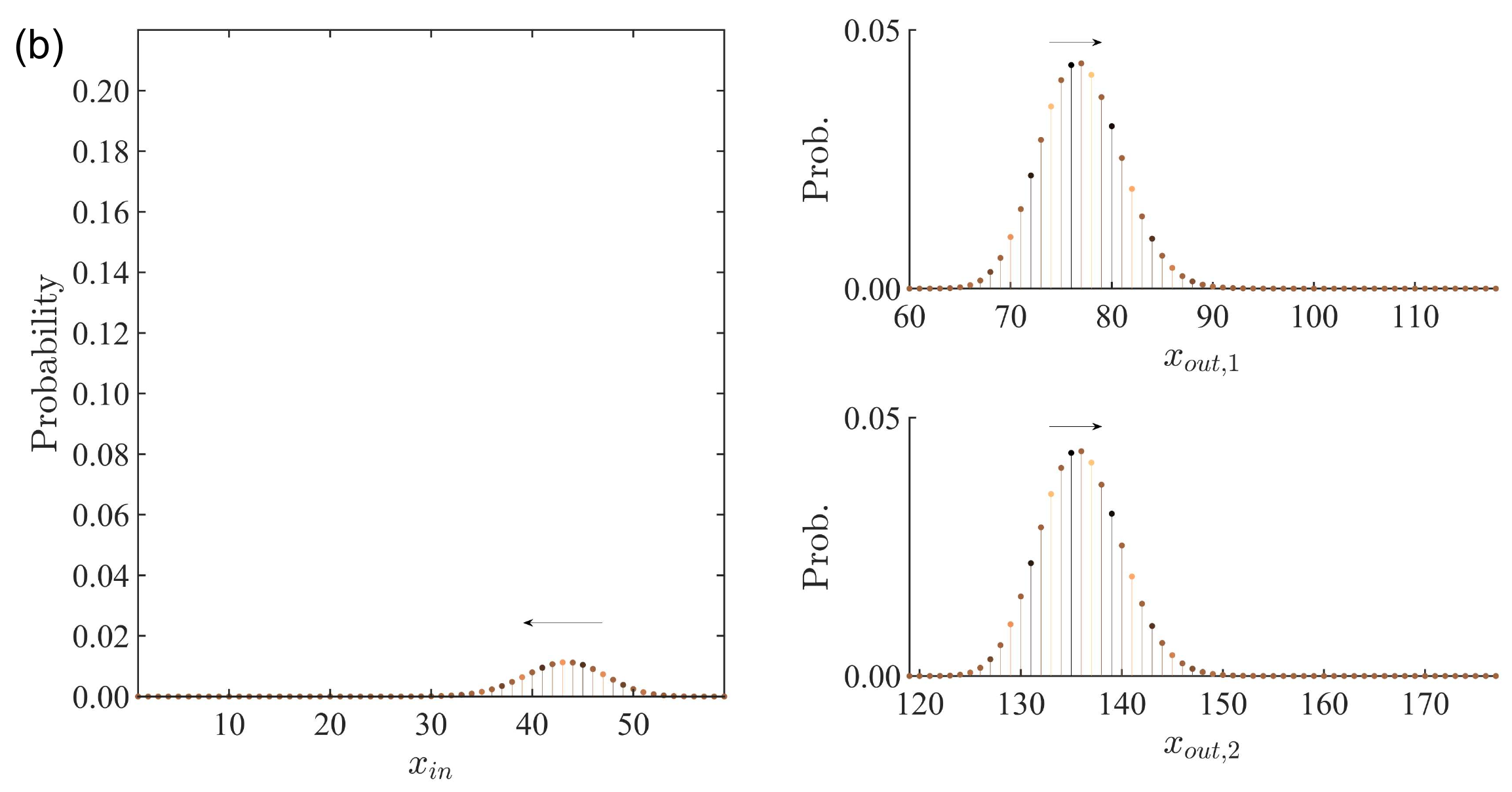}\\
\includegraphics[width=3.5in]{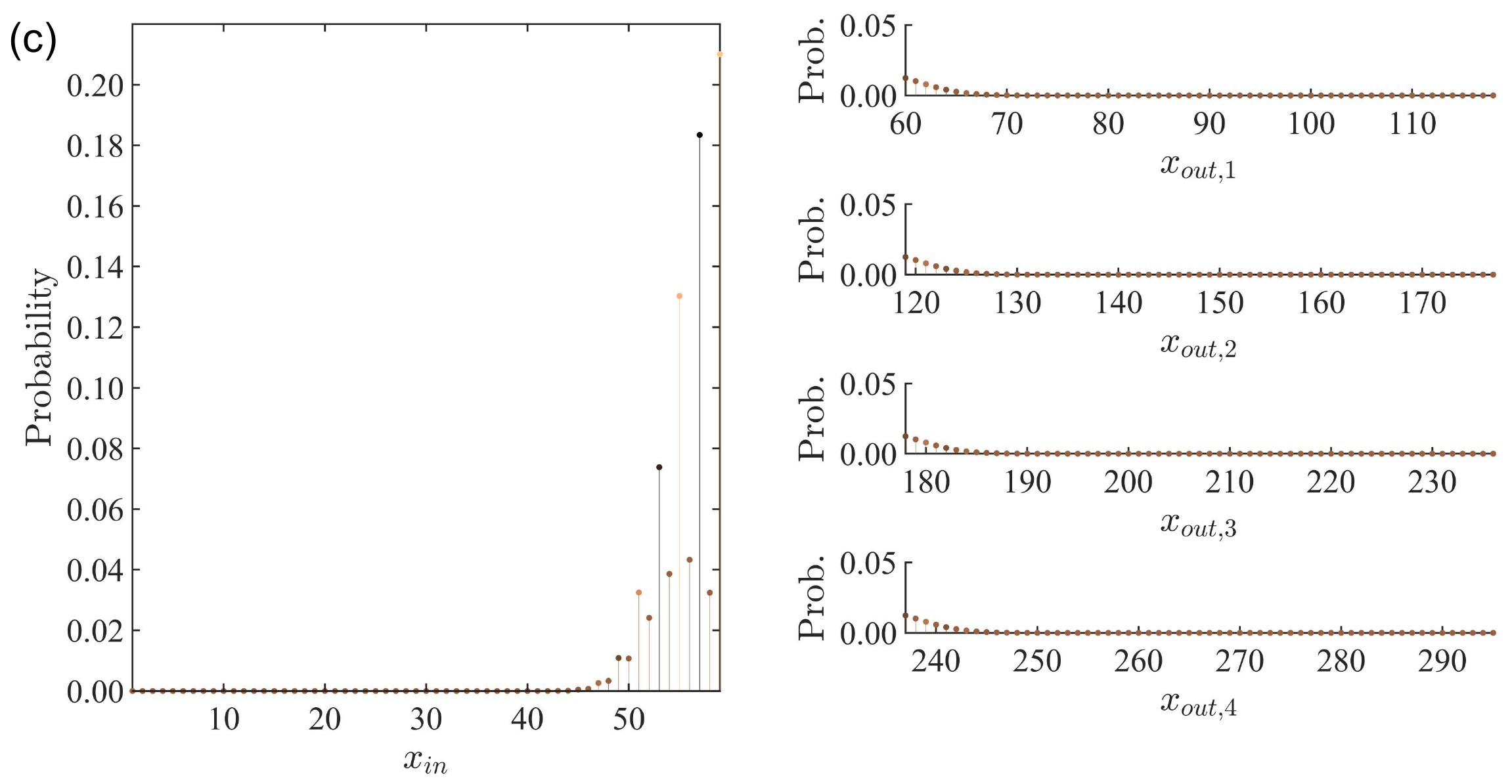}
\includegraphics[width=3.5in]{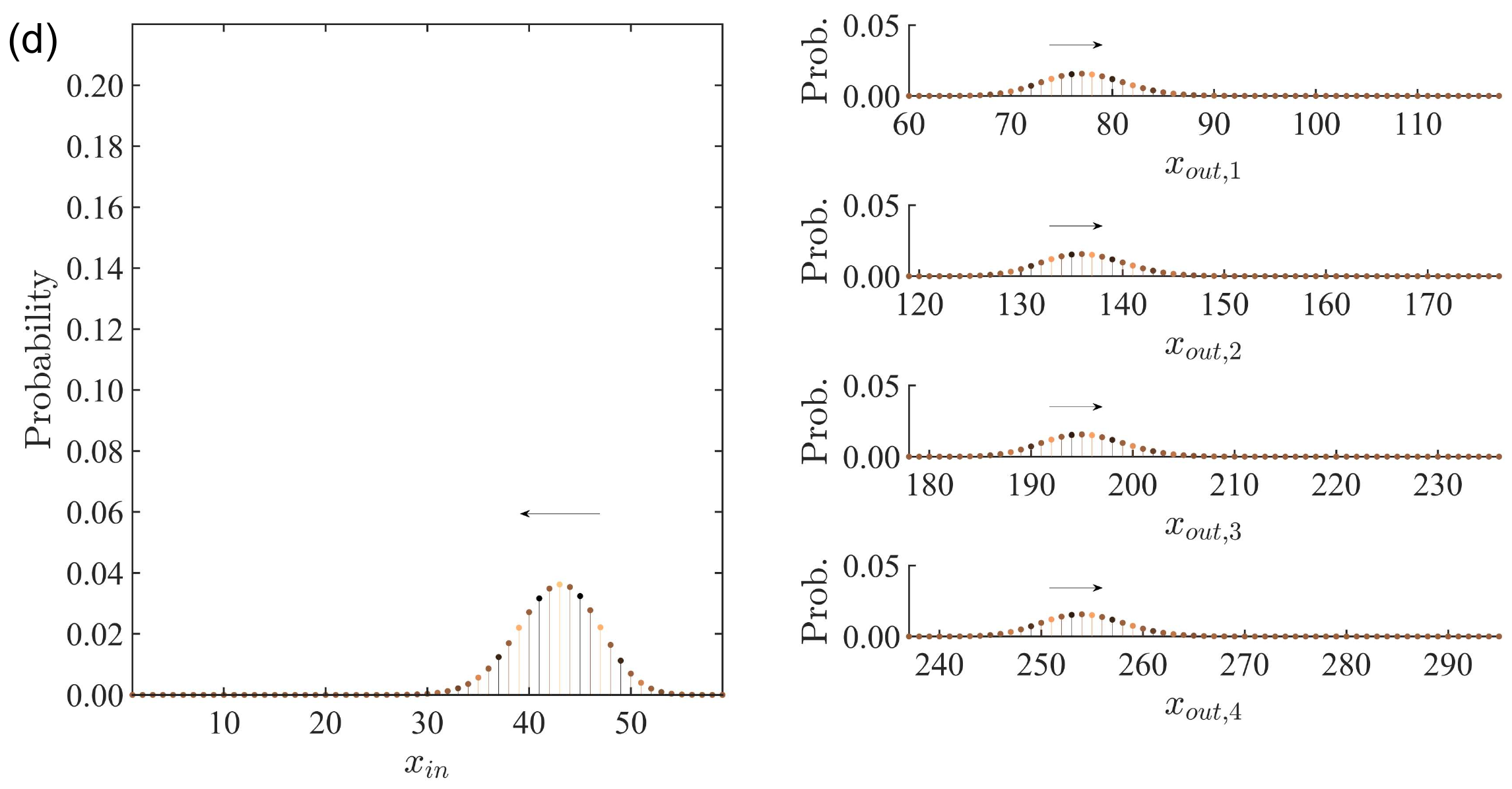}
\caption{Scattering off a junction. The top row shows a junction with two outgoing wires. The wavefunction on each wire during scattering is shown in (a). The wavefunction after scattering is shown in (b). 
The bottom row shows a junction with four outgoing wires. The wavefunction during scattering is shown in (c). The post-scattering wavefunction is shown in (d). } 
\label{fig.scatjunction}
\end{figure*}

Fig.~\ref{fig.scatjunction} shows the wavepacket impinging on a junction. After the scattering process, we are left with $(M+1)$ propagating pieces -- one moving back along the incoming wire and $M$ moving forward on outgoing wires. They represent reflected and transmitted components respectively. On each outgoing wire, we find the same transmitted component. This is consistent with our ansatz in Sec.~\ref{ssec.junctionansatz}. Reflectance is the weight of the reflected component. Fig.~\ref{fig.scatjunction} shows two cases: $M=2$ and $M=4$. As seen from the figure, reflection is stronger for $M=4$. Our simulations show that this is a general trend: reflectance increases with increasing $M$.

\begin{figure}
\includegraphics[width=3.4in]{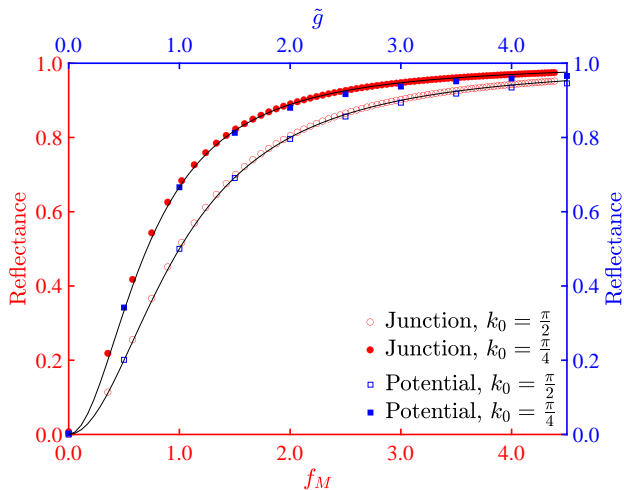}
\caption{Comparing potential and junction scattering using wavepacket dynamics. Junction scattering is represented by circles, as reflectance vs. $f_M$. Potential scattering is shown by squares, as reflectance vs. $\tilde{g}$. We show data for two choices of $k_0$, the momentum of the incoming wavepacket. Open circles/squares correspond to $k_0 = \pi/2$ while solid markers correspond to $k_0 = \pi/4$. The curves plotted are $R(x)$ vs. $x$ (see text) for the two values of $k_0$. 
 } 
\label{fig.scatpotjunc}
\end{figure}

We compare the reflectance from potentials and junctions in Fig.~\ref{fig.scatpotjunc}. In line with the equivalence discussed in Sec.~\ref{sec.equivalence} above, we consider $\tilde{g}$ and $f_M$ to be independent variables. They serve as dimensionless measures of potential strength/junction connectivity. They are plotted along the X axis in the figure. The reflectance (weight of reflected component) is the dependent variable, plotted along the Y axis. For a given value of the incoming momentum ($k_0$), the potential-reflectance and junction-reflectance data collapse onto the same curve. This curve is $R (x) \sim x^2 / (x^2 + \sin^2 k_0)$, where $x$ is $\tilde{g}$ or $f_M$. This is precisely the reflectance function found in Eqs.~\ref{eq.refg} and \ref{eq.refM}. The figure shows reflectance data for $k_0 = \pi/4$ and $\pi/2$ as well as $R(x)$ curves for the two momenta. Data points for each $k_0$ value collapse onto the corresponding $R(x)$ curve. This provides numerical verification for the arguments of Sec.~\ref{sec.equivalence}, with each junction producing the same reflectance as its equivalent potential. Fig.~\ref{fig.scatpotjunc} also shows that reflectance  increases monotonically with $\tilde{g}$ or $f_M$.

\section{Scattering in two-dimensions}
We have established an equivalence between junctions and potentials in one-dimensional wires. It is natural to ask if an equivalence also exists in higher dimensions. We consider the analogous problem in two-dimensions. On the one hand, we consider a single two-dimensional sheet with a localized (delta-function-like) potential. On the other, we consider multiple sheets that meet at a common point. This point of intersection serves as a scattering centre. We restrict our attention to point-like rather than line-like intersections. The latter case effectively reduces to a one-dimensional problem as momentum parallel to the intersection is conserved. 

In this two-dimensional setting, the `reflected' component takes the form of a scattered wave that propagates outward from the scattering centre. Simple ans\"atze such as Eqs.~\ref{eq.potansatz} and \ref{eq.Mansatz} cannot be used. For example, at long wavelengths, we may expect scattered waves to have circular symmetry with amplitudes decaying as $1/\sqrt{r}$\cite{Adhikari1986}. At the scale of the lattice, an analytic form is not known for the scattered wavefront. 
We therefore restrict ourselves to a numerical wavepacket-based approach.

In the case of potential scattering, we have a single sheet. We describe this sheet as an $L$x$L$ grid, indexed by $x$ and $y$ coordinates that each range from $-L/2$ to $+L/2$. An attractive local potential of strength $g$ is placed at the origin. 
A two-dimensional Gaussian state is initialized in one of the quadrants, 
\bea
\nn \psi_{j,initial}^{2D} = \frac{1}{\sqrt{2\pi \alpha^2}} \exp\Big\{ 
-\frac{1}{4} \frac{(x-x_0)^2  +(y-y_0)^2 }{\alpha^2} \Big\} \\
\times  \exp\Big\{  i k_{0,x}(x-x_0) + i  k_{0,y}(y-y_0) 
\Big\}. ~~
\label{eq.2dGaussian}
\eea 
This wavefunction is peaked at $(x_0,y_0)$. We choose $x_0 = y_0<0$, with the initial peak lying along the $x=y$ line. 
The momenta, $k_{0,x}$ and $k_{0,y}$, are chosen so that the initial velocity points towards the potential. The spread of the wavefunction ($\alpha$) is chosen to be  less than the initial distance to the origin.

\begin{figure*}
\includegraphics[width=7in]{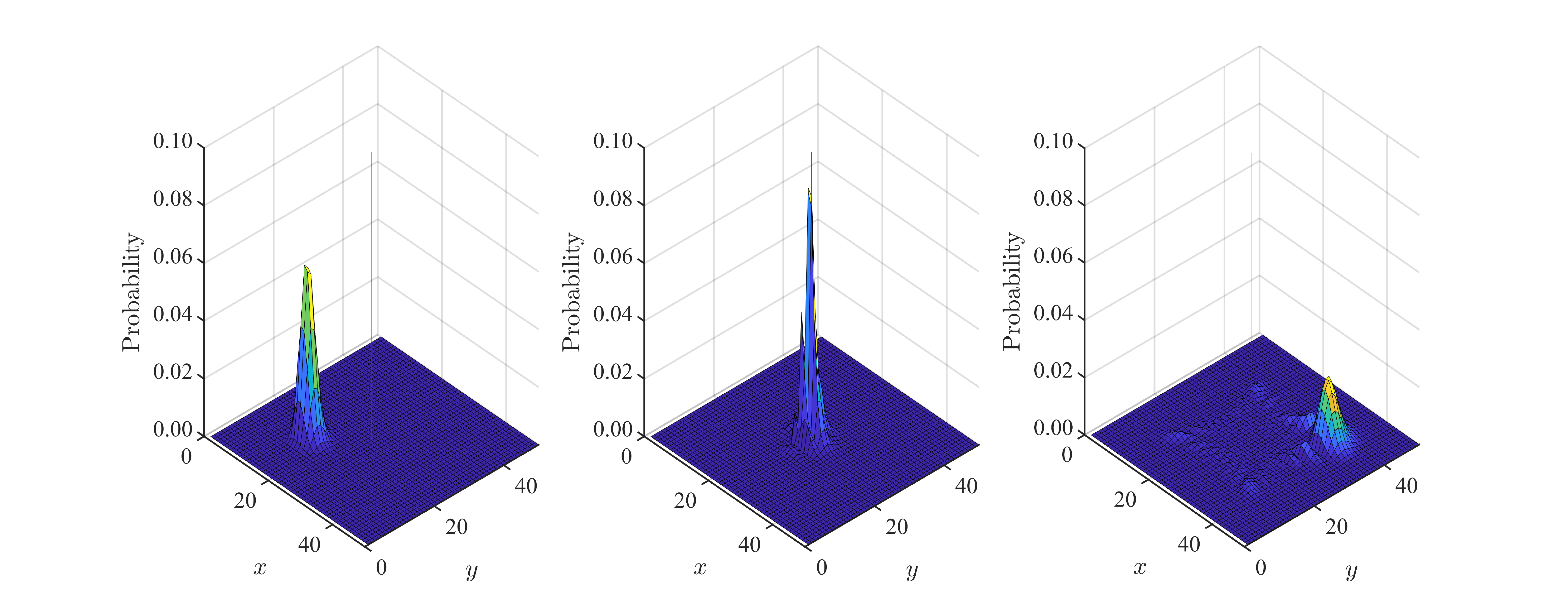}
\caption{Scattering off a potential in two dimensions. An incoming wavepacket scatters off a potential located at the origin. Wavefunctions before (left), during (centre) and after (right) scattering are shown. Post-scattering, we find a scattered component that moves outward from the origin with square-like wavefront.  } 
\label{fig.2dscat}
\end{figure*}

The scattering process is shown in Fig.~\ref{fig.2dscat}. After the wavepacket scatters off the potential at the origin, the wavefunction has two parts. One part resembles a Gaussian moving along the same direction -- this is identified as the `unscattered' component. The second moves away from the origin in the form of a square-circular ripple -- this is the `scattered' component. By comparing the wavefunction in different quadrants, we may separate the scattered and unscattered components. For example, near the $x=-y$ line, we only have contributions from the scattered component. 
In analogy with reflectance in 1D, we define the `scattered weight' in 2D. This represents the fraction that has been deflected away from the initial direction of motion. We extract this quantity by measuring the weight remaining in the original quadrant after scattering has taken place. Multiplying this quantity by four yields the scattered weight.

Fig.~\ref{fig.2dscatjun} (left) plots the scattered weight as a function of potential strength. As $g$ increases, scattered weight increases monotonically. Unlike reflectance in 1D, the scattered weight does not approach unity for large $g$. This is consistent with having a point-like potential in two dimensions: a part of the wavepacket always passes through without hitting the potential.

We next consider scattering off a junction, with one incoming sheet and $M$ outgoing sheets. All sheets share a common point, which is taken to be the origin. We initialize a Gaussian wavepacket on the incoming sheet, as given by Eq.~\ref{eq.2dGaussian}. We time-evolve this state using a tight binding Hamiltonian that is analogous to Eq.~\ref{eq.tbHam}. After scattering, the resulting wavefunction has identifiable parts. One part resembles a Gaussian that continues to move in the same direction. We identify this as the `unscattered' component -- unaffected by the junction. Other parts are square-circular ripples that propagate outward from the junction. They occur on every sheet with equal amplitude. In particular, the ripple on the incoming sheet has the same weight as that on each of the outgoing sheets. We identify these ripples as the `scattered' component. We calculate the scattered weight as the total weight in all ripples. This represents the probability of the incoming particle being deflected by the junction. 

\begin{figure*}
\includegraphics[width=7in]{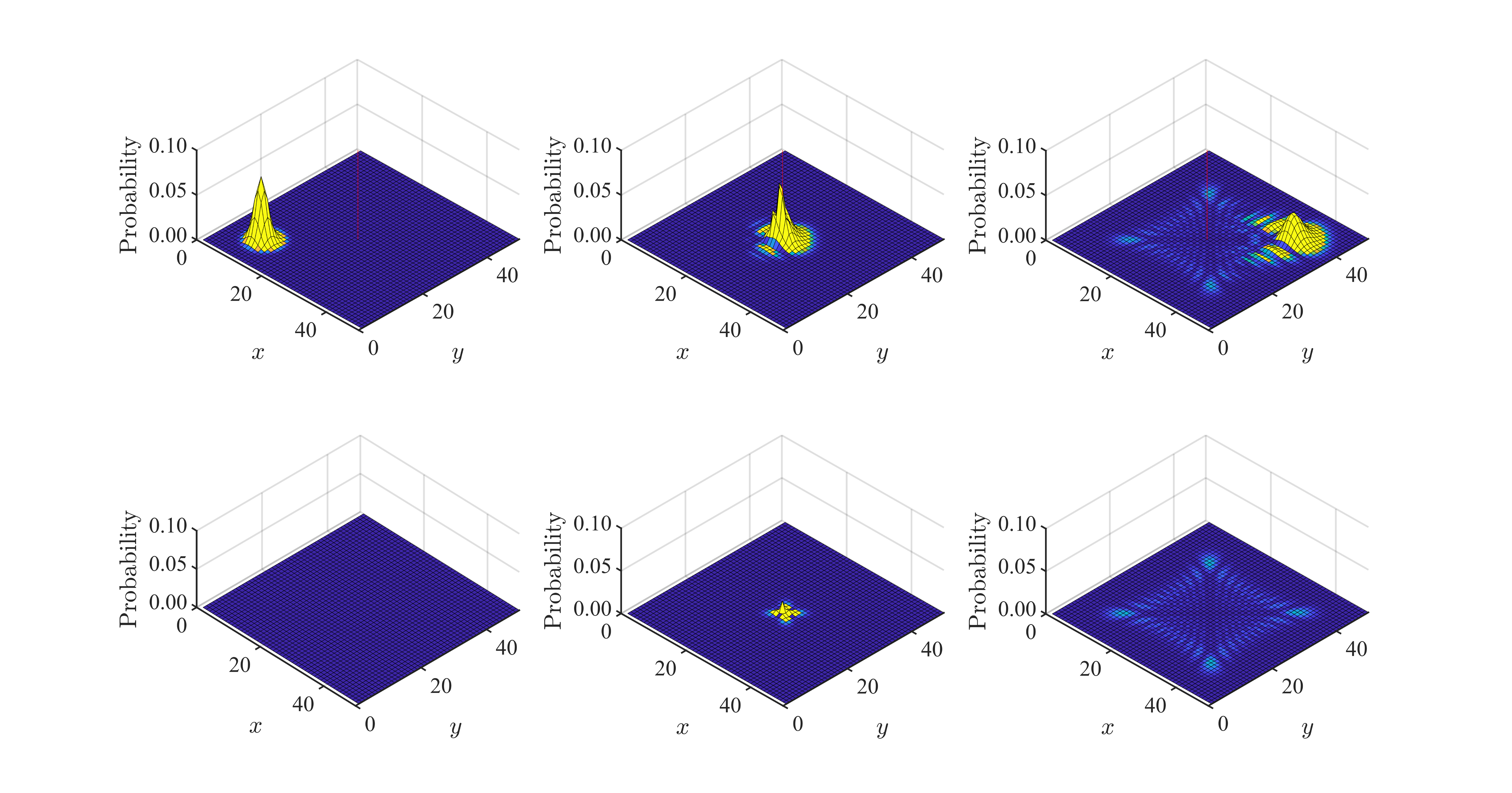}
\caption{Scattering off a junction in two dimensions. We have two sheets -- one represented in the top row and the other shown in the bottom row. The two sheets share a common point at the origin. At top left, we see an incoming wave on sheet 1 as it moves towards the origin. At this time, there is no weight on sheet 2 as shown at bottom left. Wavefunctions at the time of scattering are shown at centre-top and centre-bottom. Post scattering, the wavefunctions on each sheet are shown in the panels on the right. An unscattered component proceeds in the same direction on sheet 1. Scattered wavefronts are identical between the two sheets, proceeding outwards from the origin. } 
\label{fig.2dscatjun}
\end{figure*}

Fig.~\ref{fig.2dscatjun} (centre) shows the scattered weight as a function of $M$, the degree of connectivity of the junction. It increases monotonically with $M$ but does not approach unity.
 
By comparing Figs.~\ref{fig.2dscatjun} (left) and (centre), we may deem a junction to be equivalent to a potential if it produces the same scattered weight. For a given $M$, an equivalent potential $g_M$ can be easily found as the scattered weight increases monotonically with $M$ as well as with $g$. The obtained equivalence is shown in Fig.~\ref{fig.2dscatjun} (right) . On the X axis, we have the degree of connectivity (number of outgoing sheets) of the 2D junction. On the Y axis, we have the equivalent potential ($g$) that produces the same scattered weight. We show three data sets corresponding to three amplitudes of the incoming momentum, $k_0$. 
Unlike 1D, the equivalent potential varies with the incoming momentum.  
We compare these plots with the equivalence found in Ref.~\onlinecite{He2023} which compared 2D junctions and potentials in terms of bound state properties. As seen from the plot, the equivalent potential \`a la Ref.~\onlinecite{He2023} is different from that found by comparing scattered weights. 

The 2D equivalence shown in Fig.~\ref{fig.2dscatjun} (right) plot can be compared to the 1D result in Fig.~\ref{fig.scatpotjunc}. In both cases, as the junction's connectivity ($M$) increases, the equivalent potential increases monotonically. In 1D, the equivalence is robust and non-dispersive as it does not depend on the incoming momentum. However, in 2D, the equivalent potential shifts with momentum. This suggests that a limited notion of equivalence holds in 2D. For a coherent incoming wave with a dominant frequency, junction-scattering is equivalent to potential scattering. However, with a generic incoming wavepacket, there is no equivalence.

\begin{figure*}
\includegraphics[width=2.2in]{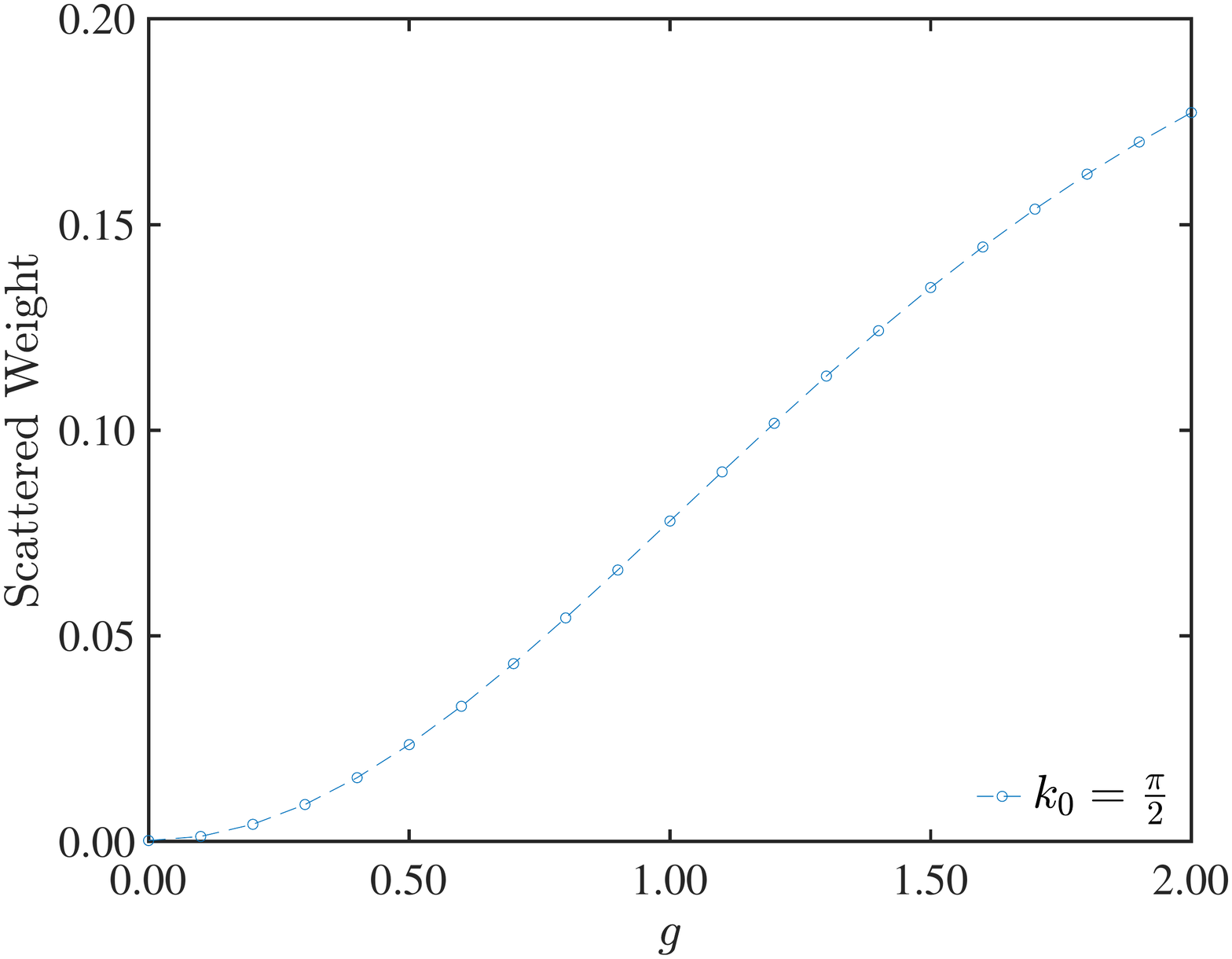}
\includegraphics[width=2.2in]{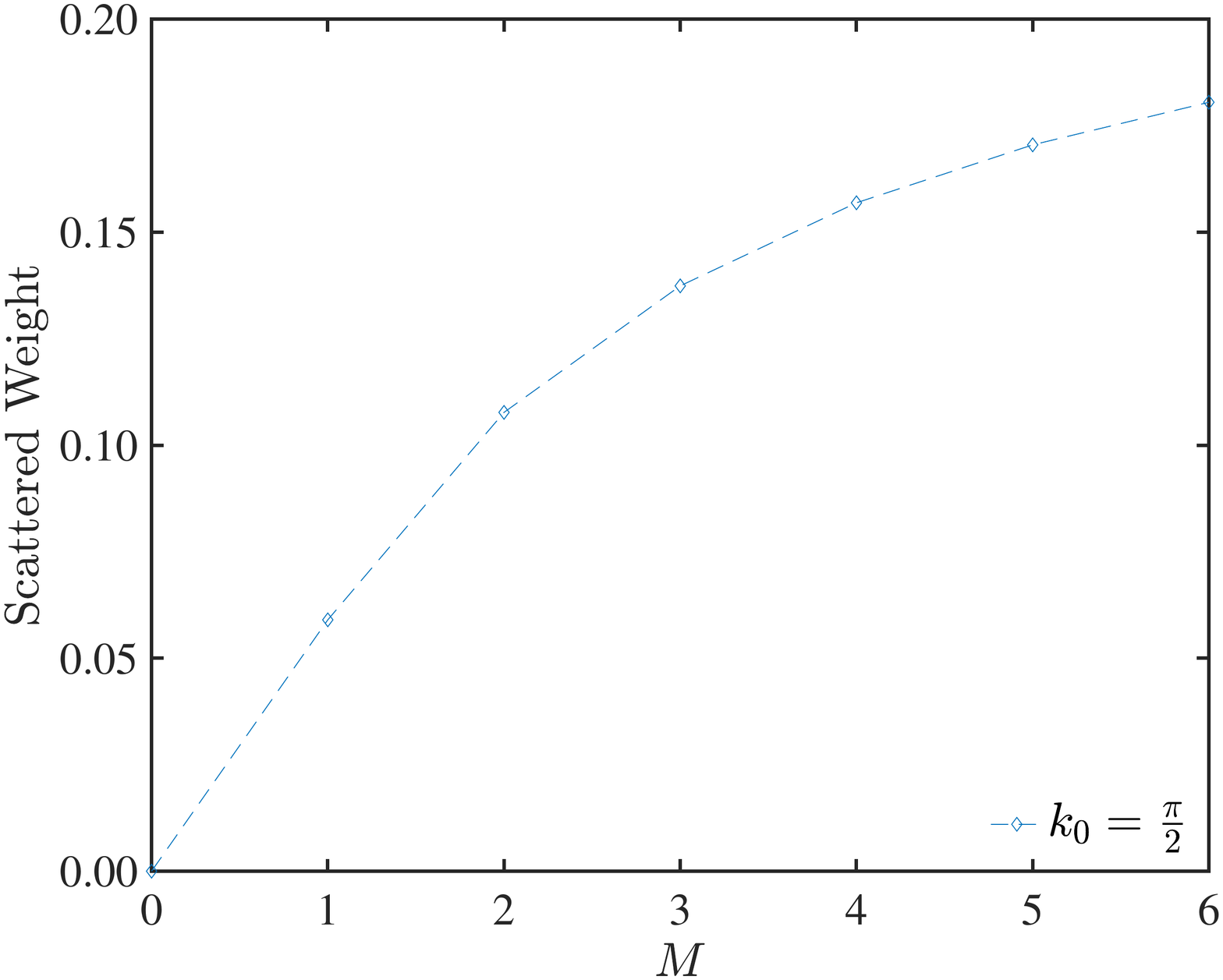}
\includegraphics[width=2.2in]{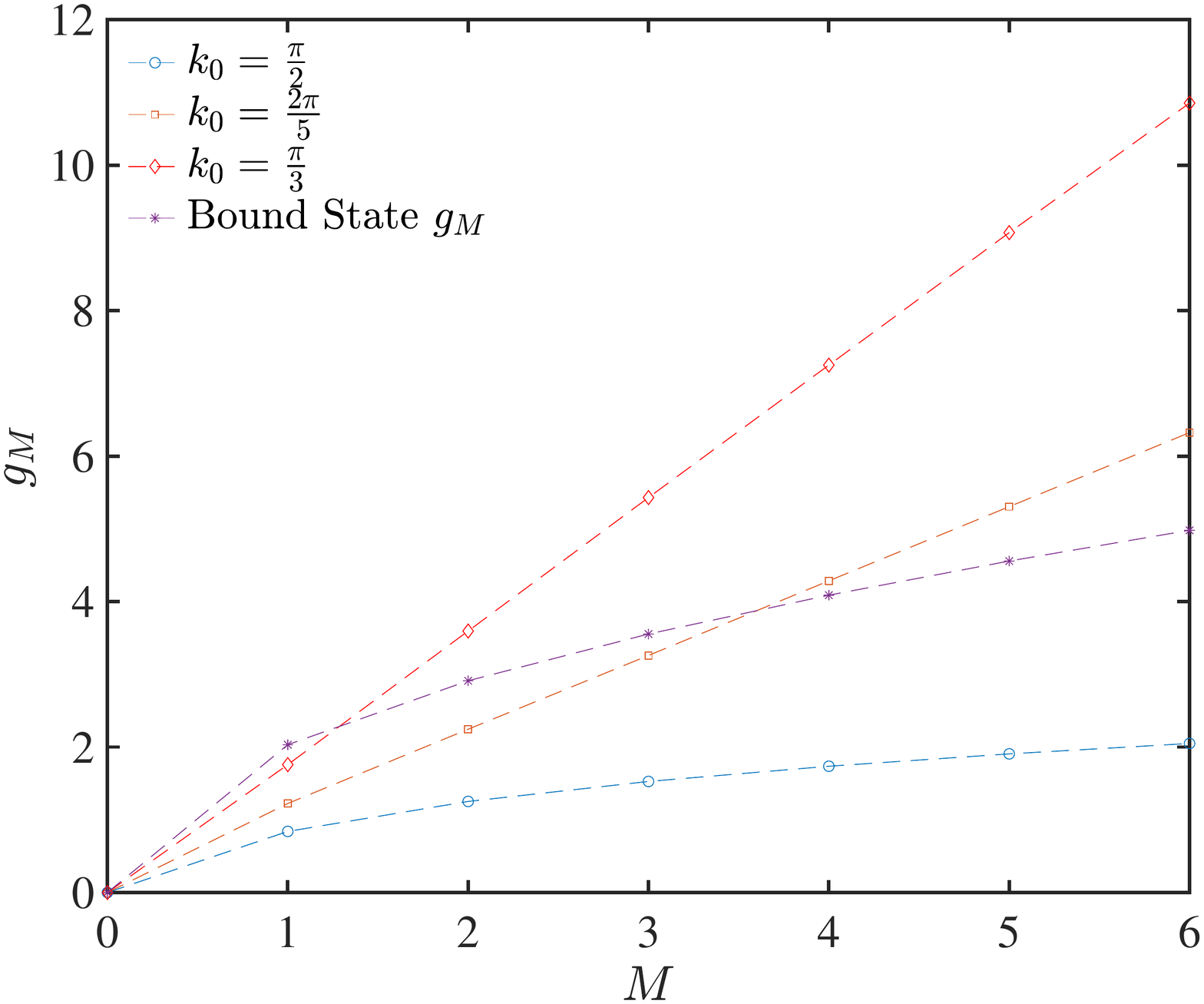}
\caption{Scattering in two dimensions. At left, we represent potential scattering with scattered weight plotted against potential strength for $k_0=\pi/2$. At the centre, we depict junction-scattering with scattered weight plotted against $M$, also for $k_0 = \pi/2$. Both functions are increase monotonically. At right, we plot the `equivalent potential' $g_M$ vs. $M$. The equivalent potential changes with incoming momentum, with data for three values of $k_0$ shown. The plot shows a fourth data set, labelled `Bound state $g_M$'. This is the equivalent potential found in Ref.~\onlinecite{He2023} -- which defines an equivalent potential as that which produces the same bound state wavefunction. } 
\label{fig.2dscatjun}
\end{figure*}

\section{Discussion}
Our central result is that a junction is quantitatively equivalent to a potential in 1D, in the sense that it produces the same reflectance. In 2D, an equivalence exists but is momentum-dependent. We have used the tight  binding approach to describe scattering. Previous studies on junction-scattering have used a continuum approach. The junction is implemented via a boundary condition\cite{Kostrykin1999} or an ersatz inter-wire coupling\cite{Tokuno2008}. The continuum and tight binding approaches have been compared in previous studies\cite{Kowal1990,Aharony2009,Soori2023}. One of our results (reflectance of a junction scales as $\sim 1/M$) agrees with unitarity-based constraints derived using a continuum description of a star graph\cite{Gratus1994}. This suggests an interesting question for future studies: does the junction-potential equivalence hold in the continuum?

Transmission through quantum networks is an issue of wide interest\cite{Kottos1999,Hennig1999,Vidal2000,Naud2001,Yang2007,Almeida2013,Caudrelier2013,Wang2022}. 
Abstract networks with fractal geometries have also evoked interest for their transmission properties\cite{Bell1994,Lin2007,Monthus2011,Xu2021}.
As networks are constructed with junctions as building blocks, our results may hold relevance. 
Our results suggest that the degree of connectivity at each junction plays a strong role. For example, an $n$-Cayley tree is built from junctions where $n$ wires meet\cite{Cayley}. Our results suggest that transmission through such as network will fall as $n$ increases.

Apart from physical networks, junction transport can play a role in abstract spaces. One example is the phenomenon of Fermi surface oscillations in metals (e.g., de Haas van Alphen effect)\cite{Ashcroft_book}. In the semiclassical limit, it is understood as the motion of electronic wavepackets on the Fermi surface. In materials such as highly-doped graphene\cite{Nikolaev2021}, multiple Fermi surfaces intersect at junctions. Fermi surface oscillations must then be understood in terms of junction-scattering. Our results for scattering amplitudes may be of use here.

\acknowledgments
We thank Diptiman Sen, Kirill Samokhin, Abhiram Soori and Eric He for insightful discussions. This work was supported by a Discovery Grant 2022-05240 from the Natural Sciences and Engineering Research Council of Canada.

\appendix
\section{Scattering coefficients from tight binding}

We first consider scattering off a potential in one dimension as described in Sec.~\ref{sec.potential}. We demand that the ansatz of Eq.~\ref{eq.potansatz} must be an eigenstate of the Hamiltonian with eigenvalue $E$. On sites that are not connected to the potential ($j < -1$ and $j > 1$), the eigenstate condition yields 
\bea
E = -2t \cos k,
\eea 
which fixes the energy eigenvalue. The eigenstate condition on sites connected to the potential ($j=\pm 1$) yields a continuity relation,
\bea
A+ B = C = D.
\eea
 Precisely at the potential site, we obtain
 \bea
 -t \big[
A e^{-i k } + B e^{+ i k }
+ D e^{ik} \big] - g C = E C.
 \eea
From these relations, the reflection coefficient $C/A$ and the transmission coefficient $D/A$ can be found. This leads to the expressions in Eq.~\ref{eq.coeffs}.

We next consider scattering off a junction, described in Sec.~\ref{sec.junction}. The ansatz in Eq.~\ref{eq.Mansatz} is to be an eigenstate of the Hamiltonian with energy eigenvalue $E$. Away from the junction ($\vert j \vert > 1$), the eigenstate condition reduces to $E = -2t\cos k$. In the immediate neighbourhood of the junction ($j = \pm 1$), the eigenstate condition yields a continuity equation, $A+B = C = D$. At the junction itself ($j=0$), the eigenstate condition yields 
\bea
- t \big[
A e^{- i k}  +  Be^{ i k} + M D e^{ik} 
\big] =  EC.
\eea
From these relations, we obtain the expressions for reflectance and transmission in Eq.~\ref{eq.MRT}.

\bibliographystyle{apsrev4-1} 
\bibliography{singularity}
\end{document}